\documentclass[%
 reprint,
 amsmath,amssymb,
 aps,
]{revtex4-1}

\usepackage{xcolor,graphicx,float}
\usepackage{subfigure}
\usepackage{dcolumn}
\usepackage{bm}
\usepackage{url}

\begin{document}


\title{Spatial search for a general multi-vertex state on graph by continuous-time quantum walks }
\thanks{A footnote to the article title}%

\author{Xi Li$^1$}
\author{Hanwu Chen$^{1,3}$}%
 \email{hw_chen@seu.edu.cn}
\author{Yue Ruan$^2$}
\author{Zhihao Liu$^{1,3}$}
\author{Mengke Xu$^1$}
\author{Jianing Tan$^1$}

\affiliation{%
 $^1$School of Computer Science and Engineering, Southeast University, Nanjing 210096, China \\
 $^2$School of Computer Science and Technology, Anhui University of Technology, Maanshan 243005,China\\
 $^3$Key Laboratory of Computer Network  and Information Integration (Southeast University), Ministry of Education, Nanjing 211189, China.
}%

\begin{abstract}
	In this work, we consider the spatial search for a general marked state on graphs by continuous time quantum walks. As a simplest case, we compute the amplitude expression of the search for the multi-vertex uniform superposition state on hypercube, and find that the spatial search algorithm is optimal for the two-vertex uniform state. However, on general graphs, a common formula can't be obtained for searching a general non-uniform superposition state. Fortunately, a Laplacian spectrum condition which determines whether the associated graph could be appropriate for performing the optimal spatial search is presented. The condition implies that if the proportion of the maximum and the non-zero minimum Laplacian eigenvalues is less or equal to $1+1/\sqrt 2$, then the spatial search is optimal for any general state. At last, we apply this condition to three kind graphs, the induced complete graph, the strongly regular graph and the regular complete multi-partite graphs. By the condition, one can conclude that these graphs will become suitable for optimal search with properly setting their graph parameters.
\end{abstract}

\maketitle

\section{\label{sec:level1}Introduction}
The study of quantum search algorithms began with Grover's study \cite{grover1997quantum}. With Grover's algorithm, one can find a marked state in an unorganized database of the size $N$ in ${\rm O}\left( {\sqrt N } \right)$ time. And Grover's algorithm is proved optimal in quantum circumstance\cite{zalka1999grover}. The success of Grover algorithm has inspired researchers to seek out quantum algorithms that have more excellent performance than classical ones, such as quantum structure search algorithm\cite{Cerf1998Nested,Roland2003Adiabatic} and Adiabatic
Quantum algorithm\cite{Albash2018Adiabatic}. One of these algorithm, called continuous time quantum walk(CTQW), is proposed by Farhi and Gutmann to solve the decision problem\cite{Farhi1998Quantum}. The CTQW can be applied in lots of situations\cite{Christandl2003Perfect,M2011Continuous,Kulvelis2015Universality,Galiceanu2016dendrimer,Qiang2012An,Gamble2010Two} and it is an universal quantum algorithm\cite{Childs2009Universal}. By using CTQW, Childs and Goldstone presented quantum spacial algorithm on periodic lattice\cite{childs2004spatial}. Afterward, search algorithms based on the CTQW have been investigated. Past research has mainly focused on the time complexity of special graphs such as complete graphs and hypercubes. These studies have shown that various types of graph are suitable for quantum searches. Janmark and Meyer\cite{janmark2014global} have shown that global symmetry is unnecessary for a fast quantum search and that a strongly regular graph can be used for fast searching. Subsequently, Meyer and Wong utilized a parallel computation method (theory of degenerate perturbation) and concluded that connectivity is a poor indicator of a fast quantum search. These studies revealed partial correlations between the graph structures and the search performance.

For many special graphs, the CTQW search algorithm can detect the marked state in time ${\rm O}\left( {\sqrt N } \right)$. 
The search algorithm is optimal on the associated graph if only ${\rm O}\left( {\sqrt N } \right)$ time is required for finding the marked state. With slightly changing the original search algorithm, by using the adjacent matrix to replace the Laplacian matrix, Chakraborty and Leonardo\cite{chakraborty2016spatial} demonstrated that a random graph of $n$ vertices where each edge exists with probability $p$, search by CTQW is almost surely optimal as long as $p \ge log^{3/2}N/N$.
However, when the associated graph isn't regular, then the adjacent matrices and Laplacian matrices will produce different effects. And the Laplacian matrix has significant applications in other areas. Hence, we adopt the original algorithm version. 

Besides the single vertex solution search, the multiple solutions search has likewise been considered\cite{Nielsen2011Quantum,Williams2011Explorations,Wong2016Spatial}. In this work, we study multiple solutions search. We firstly compute the multiple vertex uniform superposition state search performance on hypercube. Then we generalize the uniform state to a non-uniform state, we want to obtain an optimal seaarch condition of the spectrum of associated graph for any target state. By default, we require the overlap between the initial state and the target state is bigger than $1/ \sqrt N$ .  

We organize this work as follows. In the second section, we review the spacial search by CTQW on a general graph. In the third and fourth section, we present the multi-solutions search on the hypergraph. In the fifth section, we will give the optimal spectrum condition for finding a general marked state. In the sixth section, we will give three kinds of graphs on which we can perform optimal quantum search, they are induced conmplete graph, strongly regular graph and regular complete multipartite graph. 

\section{\label{sec:level1}Review of Spatial Search by CTQW}
The CTQW is defined on a graph. Let $G\left( {V(G),E(G)} \right)$ be a connected graph with edge set $E(G)$ and vertex set $V(G) = \left\{ {1, \ldots N} \right\}$, where the vertex set corresponds to the computational basis of a $N$-dimensional Hilbert space and is denoted by $\left\{ {\left| 1 \right\rangle , \ldots ,\left| N \right\rangle } \right\}$,
The system evolves with the Hamiltonian
\begin{equation}\label{s1.1}
H =  -\gamma L - \left| w \right\rangle \left\langle w \right|,
\end{equation}
where \emph{L=A-D} is the Laplacian matrix of the graph, \emph{A} and \emph{D} are adjacency matrix and degree matrix respectively. $ \gamma$ is the jumping rate from a vertex to its adjacent vertex, $\left| w \right\rangle $ is the marked state and $- \left| w \right\rangle \left\langle w \right|$  is the oracle. The initial state is a uniform superposition state:
\begin{equation}
\left| s \right\rangle  = \frac{1}{\sqrt N}\sum\limits_{i = 1}^N {\left| i \right\rangle } .
\end{equation}
Let the spectrum of \emph{-L} be $\left\{ {{\lambda _1}, \ldots ,{\lambda _N}} \right\}$ in a non-incremental arrangement, and ${\lambda _1} \ge  \cdots  \ge {\lambda _{N - 1}} > {\lambda _N} = 0$; the corresponding eigenvectors are $\left\{ {\left| {{{\bf{\lambda }}_{\bf{1}}}} \right\rangle {\bf{,}} \ldots {\bf{,}}\left| {{{\bf{\lambda }}_{\bf{N}}}} \right\rangle } \right\}$. Note that $\left| s \right\rangle  = \left|  {{\lambda _N}} \right\rangle$ is the eigenvector belonging to the minimum eigenvalue $0$. We let the eigenvalues set of the Hamiltonian $\emph{H}$ be $\left\{ {{\mu _1}, \ldots ,{\mu _N}} \right\}$. The eigenvectors set belonging to the eigenvalues is $\left\{ {\left| {{\mu _1}} \right\rangle , \ldots ,\left| {{\mu _N }} \right\rangle } \right\}$. The eigenequation is
\begin{equation}\label{s1.2}
H\left| {{\mu _\kappa }} \right\rangle  = {\mu _\kappa }\left| {{\mu _\kappa }} \right\rangle.
\end{equation}
At time $t$, the probability amplitude of detecting the marked state is:
\begin{equation}\label{s1.3}
\left\langle w \right|\mathop e\nolimits^{ - iHt} \left| s \right\rangle  = \sum\limits_k {\left\langle w \right.\left| {{\mu _k}} \right\rangle } \left\langle {{\mu _k}} \right.\left| s \right\rangle \mathop e\nolimits^{ - i{\mu_k}t}.
\end{equation}
Define ${P_k} = \left\langle {w\left| {{\lambda _k}} \right\rangle } \right.$ and $ f\left( \mu  \right) = \sum\limits_k {\frac{{P_k^2}}{{\gamma {\lambda _k} - \mu }}}$; using the method of Childs and Goldstone \cite{childs2004spatial}, we have:
\begin{equation}\label{s1.4}
\left\langle w \right|\mathop e\nolimits^{ - iHt} \left| s \right\rangle  =
-P_N \sum\limits_k {\frac{{\mathop e\nolimits^{ - i{\mu _k}t} }}{{{\mu _k}{f^\prime}({\mu _k})}}}.
\end{equation}
The estimation of expression (\ref{s1.4}) is:
\begin{equation}\label{s1.50}
\left| {\left\langle w \right|\mathop e\nolimits^{ - iHt} \left| s \right\rangle } \right|  \approx P_N \left| {\frac{{2\sin \left( {{\mu _1}t} \right)}}{{{\mu _1}{f^\prime}\left( {{\mu _1}} \right)}}} \right|  .
\end{equation}
This can be written as 
\begin{equation}\label{s1.5}
\left| {\left\langle w \right|\mathop e\nolimits^{ - iHt} \left| s \right\rangle } \right| \approx  \frac{\gamma }{\beta }\left| {\sin \left( {\frac{\gamma P_N}{{\beta }}t} \right)} \right|.
\end{equation}
Where $\gamma$ and $\beta$ are two parameter derived from the eigenvalues and eigenvectors:
\begin{equation}\label{s1.6}
\left\{ {\begin{array}{*{20}{c}}
	{\gamma {\rm{ = }}\sum\limits_{k \ne N} {\frac{{P_k^2}}{{{\lambda _k}}}} }\\
	\\
	{\beta {\rm{ = }}\sqrt {\sum\limits_{k \ne N} {\frac{{P_k^2}}{{\lambda _k^2}}} } }
	\end{array}} \right.
\end{equation}
With the sinusoidal form of equation(\ref{s1.5}), we can determine the optimal search time and the corresponding amplitude.
One can observe that when $T = \frac{{\pi \beta }}{{2\gamma P_N}}$ the amplitude of the marked state is maximum. By default, the overlap between the marked state and the initial state is over than $\frac{1}{\sqrt N}$. If $\gamma / \beta \in \left( 1/{\sqrt{2}} , 1\right )$, $T = {\rm O}\left( {\sqrt N } \right)$ is also satisfied. Hence the optimal condition for a graph is as follows:
\begin{equation} \label{s1.7}
{\frac{\gamma }{\beta } \in \left( {\frac{1}{\sqrt{2}},1} \right)}.
\end{equation}
And The ${\gamma  \mathord{\left/
		{\vphantom {\gamma  \beta }} \right.
		\kern-\nulldelimiterspace} \beta }
\approx 1$ is the ideal case. Using the definitions in equation (\ref{s1.6}), we have
\begin{equation} \label{s1.8}
{\frac{1}{\sqrt 2} < \frac{{\sum\limits_{k \ne n} {\frac{{P_k^2}}{{{\lambda _k}}}} }}{{\sqrt {\sum\limits_{k \ne n} {\frac{{P_k^2}}{{\lambda _k^2}}} } }} < 1}.
\end{equation}
The (\ref{s1.8}) describes that what kind of graphs suitable for performing optimal spatial search by CTQW. The Laplacian eigenvalues and eigenvectors need to be calculated for the associated graph. In general case, the calculation of eigenvectors is a troublesome, fortunately, for some graph that isn't so intricate, such as hypercube. We will compute the eigenvalues and eigenvectors of the hypercube, and analyze the performance of the search algorithm in the next section. 

\section{\label{sec:level1}Two Solutions Search on The Hypercube Graph}
Spatial search by CTQW is optimal on hypercube graph single vertex state\cite{childs2004spatial}. In this section, we will show that the search is equally optimal for searching a two-vertex uniform superposition state on hypercube. 
Before that, we will introduce the Cartesian product of graphs and give the Laplacian spectrum relationship between the result graph and the original graphs.

For two given graph $G$ and $H$, their Cartesian product is the graph $G \square H$ whose vertex set is $V(G) \times V(H)$ and whose edge set is the set of all pairs $(u_1, v_1)(u_2, v_2)$ such that either $u_1u_2 \in E(G)$ and $v_1 = v_2$, or $v_1v_2 \in E(H)$ and $u_1 = u_2$. The follow lemma \cite{Fiedler1973Algebraic,merris1994laplacian} tells us how to compute the Laplacian spectrum of the Cartesian product graph. 

$\bm{Lemma}$. Let $G$ and $H$ be graphs on $n_1$ and $n_2$ vertices respectively. Then the eigenvalues of $G \square H$ are all possible sums ${\lambda _j}(G) + {\lambda _k}(H)$, $1\le j\le n_1$ and $1\le k\le n_2$

The hypercube $Q_N$ can be generated by the Cartesian product of $\log \left( N \right)$ two-vertices complete graphs $K_2$. Without loss of generality, we let $\log \left( N \right)$ be even number. We can represent the vertex of $Q_N$ by a binary string of length $\log \left( N \right)$, e.g, $\left| 0 \right>=(0,\ldots,0)$ represents the original node of $Q_N$. For a given string, we can obtain its corresponding quantum state by taking tensor product of each coordinate, e.g, $(1,\ldots,1)$ can be represented as $|N-1\rangle = |1\rangle \otimes \cdots \otimes |1\rangle$.

From \textbf{lemma 1}, we can obtain the eigenvalus of $Q_N$, they are
\begin{equation} \label{s2.1}
{\lambda _{{j_1}, \ldots ,{j_n}}}\left( {{Q_N}} \right) = \sum\limits_{k = 1}^2 {{\lambda _{jk}}\left( {{K_2}} \right)},
\end{equation}
where $n=\log(N)$ and corresponding eigenstates are\cite{Mieghem2011Graph} 
\begin{equation} \label{s2.2}
\left| {{\lambda _{{j_1}, \ldots ,{j_n}}}\left( {{Q_N}} \right)} \right\rangle  = \left| {{\lambda _{j1}}\left( {{K_2}} \right)} \right\rangle  \otimes  \cdots  \otimes \left| {{\lambda _{jn}}\left( {{K_2}} \right)} \right\rangle,
\end{equation}
where the $\lambda_{jk}(K_2)$, are Laplacian eigenvalues of $K_2$, equal to $0$ or $2$, and corresponding eigenstates are $\left|\lambda_{jk}(K_2)\right>=\left[ {\frac{1}{{\sqrt 2 }},\frac{1}{{\sqrt 2 }}} \right]^T$ or $\left|\lambda_{jk}(K_2)\right>=\left[ {\frac{1}{{\sqrt 2 }},\frac{-1}{{\sqrt 2 }}} \right]^T$.
We choose $\frac{1}{{\sqrt 2 }}\left( {\left| 0 \right\rangle  + \left| {N - 1} \right\rangle } \right)$ as the marked state.
Bring the eigenvalues and eigenstates into (\ref{s1.6}) to obtain 
\begin{equation} \label{s2.3}
\gamma  = \sum\limits_{{j_1} = 1}^2 {\sum\limits_{{j_2} = 1}^2 { \cdots \sum\limits_{{j_n} = 1}^2 {\frac{{\frac{1}{2}{{\left( {\prod\limits_{k = 1}^n {\left\langle 0 \right.\left| {{\lambda _{jk}}} \right\rangle }  + \prod\limits_{k = 1}^n {\left\langle 1 \right.\left| {{\lambda _{jk}}} \right\rangle } } \right)}^2}}}{{\sum\limits_{k = 1}^n {{\lambda _{jk}}\left( {{K_2}} \right)} }}} } }.
\end{equation}
Since 
\begin{equation}
\prod\limits_{k = 1}^n {\left\langle 0 \right.\left| {{\lambda _{jk}}} \right\rangle }  + \prod\limits_{k = 1}^n {\left\langle 1 \right.\left| {{\lambda _{jk}}} \right\rangle }  = \left( {1 + {{\left( { - 1} \right)}^f}} \right){\left( {\frac{1}{{\sqrt 2 }}} \right)^n},
\end{equation}
where $f$ is the number of eigenvalus $2$ for one of the eigenvalue combinations, Hence 
\begin{equation}
\gamma  = {{\frac{1}{2^{n-1}}}}\sum\limits_{l = 1}^{{\raise0.7ex\hbox{$n$} \!\mathord{\left/
			{\vphantom {n 2}}\right.\kern-\nulldelimiterspace}
		\!\lower0.7ex\hbox{$2$}}} {\left( {\begin{array}{*{20}{c}}
		n\\
		{2l}
		\end{array}} \right)\frac{1}{{4l}}},
\end{equation}
and 
\begin{equation}
\beta  = \sqrt {{{{\frac{1}{2^{n - 1}}}}}\sum\limits_{l = 1}^{{\raise0.7ex\hbox{$n$} \!\mathord{\left/
				{\vphantom {n 2}}\right.\kern-\nulldelimiterspace}
			\!\lower0.7ex\hbox{$2$}}} {\left( {\begin{array}{*{20}{c}}
			n\\
			{2l}
			\end{array}} \right)\frac{1}{{{{\left( {4l} \right)}^2}}}} }.
\end{equation}
We can represent $\gamma$ and $\beta$ as generalized hypergeometric series\cite{HypergeometricPFQ} and we can obtain the numerical value. For large $n$, $\gamma/\beta\approx1$, therefore, the quantum search is optimal on hypercube graph for the search of this special state $ \frac{1}{2}\left( {\left| 0 \right\rangle {\rm{ + }}\left| {N - 1} \right\rangle } \right)
$. For general case, if $p_1$ and $p_2$ are arbitrary two positions in hypercube, we let the target state 
\begin{equation} \label{s2.4}
\left| w \right\rangle  = \frac{1}{{\sqrt 2 }}\left( {\left| {{p_1}} \right\rangle {\rm{ + }}\left| {{p_2}} \right\rangle } \right).
\end{equation}
As the hypercube is symmetric, any vertex can be regarded as the origin. Then $\left|w\right>$ can be written as 
\begin{equation} \label{s2.5}
\left| w \right\rangle  = \frac{1}{{\sqrt 2 }}\left( {\left| {{0}} \right\rangle {\rm{ + }}\left| {{p_2}} \right\rangle } \right).
\end{equation}
We have 
\begin{equation} \label{s2.6}
\gamma  = \sum\limits_{{j_1} = 1}^2 {\sum\limits_{{j_2} = 1}^2 { \cdots \sum\limits_{{j_n} = 1}^2 {\frac{{\frac{1}{2}\frac{1}{{{2^{n - m}}}}{{\left( {\prod\limits_{l = 1}^m {\left\langle 0 \right.\left| {{\lambda _{jl}}} \right\rangle }  + \prod\limits_{k = 1}^m {\left\langle 1 \right.\left| {{\lambda _{jl}}} \right\rangle } } \right)}^2}}}{{\sum\limits_{l = 1}^n {{\lambda _{jl}}\left( {{K_2}} \right)} }}} } },
\end{equation}
where $m$ is the number of $1$ in the binary string of vertex $\left|p_2\right>$, and we set $m$ is a even number. With combination method, we have 
\begin{equation} 
\gamma  = A_1 + A_2,
\end{equation}
where 
\begin{equation}
A_1 = \frac{1}{{{2^{n-1}}}}\sum\limits_{l = 1}^{n - m} {\sum\limits_{p = 0}^{{m \mathord{\left/
				{\vphantom {m 2}} \right.
				\kern-\nulldelimiterspace} 2}} {\left( {\begin{array}{*{20}{c}}
			n-m\\
			l
			\end{array}} \right)\left( {\begin{array}{*{20}{c}}
			m\\
			{2p}
			\end{array}} \right)\frac{1}{{2\left( {l + 2p} \right)}}} },
\end{equation}
and 
\begin{equation}
A_2= \frac{1}{{{2^{n-1}}}} \sum\limits_{p = 1}^{{m \mathord{\left/
			{\vphantom {m 2}} \right.
			\kern-\nulldelimiterspace} 2}} {\left( {\begin{array}{*{20}{c}}
		m\\
		{2p}
		\end{array}} \right)\frac{1}{{4p}}} .
\end{equation}
Similarly 
\begin{equation} 
\beta^2 = B_1 + B_2,
\end{equation}
where 
\begin{equation}
B_1 = \frac{1}{{{2^{n-1}}}}\sum\limits_{l = 1}^{n - m} {\sum\limits_{p = 0}^{{m \mathord{\left/
				{\vphantom {m 2}} \right.
				\kern-\nulldelimiterspace} 2}} {\left( {\begin{array}{*{20}{c}}
			n-m\\
			l
			\end{array}} \right)\left( {\begin{array}{*{20}{c}}
			m\\
			{2p}
			\end{array}} \right)\frac{1}{{4{{\left( {l + 2p} \right)}^2}}}} },
\end{equation}
and 
\begin{equation}
B_2 = \frac{1}{{{2^{n-1}}}}\sum\limits_{p = 1}^{{m \mathord{\left/
			{\vphantom {m 2}} \right.
			\kern-\nulldelimiterspace} 2}} {\left( {\begin{array}{*{20}{c}}
		m\\
		{2p}
		\end{array}} \right)\frac{1}{{16{p^2}}}} .
\end{equation}
Hence, we can obtain the values of $\gamma$ and $\beta$ once we know the two positions by the given equation. From TABLE
$\uppercase\expandafter{\romannumeral1}$, the quantum search on htpercube by CTQW is optimal for two-vertex uniform superposion state.  
\begin{table}[!htbp]
	\caption{The search amplitude of two-vertex uniform superposition.
		The number of total positions are $16$, $m$ is the number of different positions of the two vertex.}
	\centering
	\begin{tabular}{|c|c|c|c|c|c|c|c|c|c|c|c|c|c|c|c|c|}
		\hline
		{$m$}&$1$&$2$&$3$&$4$&$5$&$6$&$7$&$8$\\ 
		\hline
		$\gamma/\beta$&0.9418&0.9374
		&0.9422&0.9448&0.9462&0.9471&0.9477&0.9481
		\\
		\hline
		{$m$}&$9$&$10$&$11$&$12$&$13$&$14$&$15$&$16$\\ 
		\hline
		$\gamma/\beta$&0.9485&0.9488&0.9491&0.9492
		&0.9494
		&0.9496
		&0.9497
		&0.9498
		\\
		\hline	
	\end{tabular}
	\\
\end{table}
\section{\label{sec:level1}Multi-Solutions Search on The Hypercub Graph}
Now we will consider multi-vertex uniform superposition state search
on hyperon $Q_N$. Let $m$ states group be as follows 
\begin{equation}
\left\{ {\begin{array}{*{20}{c}}
	{\left| {{s_1}} \right\rangle  = \left| {10, \ldots ,0} \right\rangle }\\
	{\left| {{s_2}} \right\rangle  = \left| {01, \ldots ,0} \right\rangle }\\
	\vdots \\
	{\left| {{s_m}} \right\rangle  = \left| {0, \ldots ,1, \ldots ,0} \right\rangle }
	\end{array}} \right.
\end{equation}
and $\left| w \right\rangle  = \frac{1}{{\sqrt m }}\sum\limits_{j = 1}^m {{s_j}}$. Using the (\ref{s1.6}), we have 
\begin{equation} \label{s3.1}
\gamma  = \sum\limits_{{j_1} = 1}^2 { \cdots \sum\limits_{{j_n} = 1}^2 {\frac{{{{\left( {\frac{1}{{\sqrt m }}{{\left( {\frac{1}{{\sqrt 2 }}} \right)}^{n - m}}\sum\limits_{l = 1}^m {\left\langle {{\lambda _{jl}}} \right.\left| 1 \right\rangle } } \right)}^2}}}{{\sum\limits_{l = 1}^n {{\lambda _{jl}}\left( {{K_2}} \right)} }}} } .
\end{equation}
Simplify (\ref{s3.1}) to obtain 
\begin{equation}
\gamma  = \frac{1}{{{2^n}}}\left( {{A_1} + {A_2}} \right),
\end{equation}
where 
\begin{equation}
{A_1}{\rm{ = }}\mathop \sum \limits_{l = 1}^{n - m} \mathop \sum \limits_{p = 0}^m \left( {\begin{array}{*{20}{c}}
	{n - m}\\
	l
	\end{array}} \right)\left( {\begin{array}{*{20}{c}}
	m\\
	p
	\end{array}} \right)\frac{{{{\left( {m - 2l} \right)}^2}}}{{2m\left( {l + p} \right)}},
\end{equation}
and 
\begin{equation}
{A_2} = \mathop \sum \limits_{l = 1}^m \left( {\begin{array}{*{20}{c}}
	m\\
	l
	\end{array}} \right)\frac{{{{\left( {m - 2l} \right)}^2}}}{{2ml}}.
\end{equation}
Similarly, we have 
\begin{equation}
{\beta ^2} = \frac{1}{{{2^n}}}\left( {{B_1} + {B_2}} \right),
\end{equation}
where 
\begin{equation}
{B_1} = \sum\limits_{l = 1}^{n - m} {\sum\limits_{p = 0}^m {\left( {\begin{array}{*{20}{c}}
			{n - m}\\
			l
			\end{array}} \right)\left( {\begin{array}{*{20}{c}}
			m\\
			p
			\end{array}} \right)\frac{{{{\left( {m - 2l} \right)}^2}}}{{m{{\left( {2\left( {l + p} \right)} \right)}^2}}}} } ,
\end{equation}
and 
\begin{equation}
{B_2} = \sum\limits_{l = 1}^k {\left( {\begin{array}{*{20}{c}}
		m\\
		l
		\end{array}} \right)\frac{{{{\left( {m - 2l} \right)}^2}}}{{m{{\left( {2l} \right)}^2}}}} .
\end{equation}
Through numerical methods, we have $\gamma / \beta \approx 1$, this means that hypercub is optimal for search this type of superposition states. 

For general cases, it's hard to obtain a common expression, since the number of different combinations of position is $2^n$ and the number of different eigenvalue combinations is also $2^n$. However, we can obtain search amplitude by using the result of the single vertex and two-vertex search. We let $m=3$ first, the format of a three positions uniform superposition is 
\begin{equation} \label{s3.2}
\left| w \right\rangle  = \frac{1}{{\sqrt 3 }}\left( {\left| p_1 \right\rangle  + \left| {{p_2}} \right\rangle {\rm{ + }}\left| {{p_3}} \right\rangle } \right).
\end{equation}
Use the original definition of $\gamma$ 
\begin{equation}\label{s3.3}
\gamma  = \sum\limits_{k \ne N} {\frac{{{{\left| {\left\langle w \right|\left. {{\lambda _k}} \right\rangle } \right|}^2}}}{{{\lambda _k}}}}.  
\end{equation}
We take the $\left|w\right>$ into to obtain
\begin{equation}\label{s3.4}
\gamma  = \frac{1}{3}\left( {{\gamma _{12}} + {\gamma _{13}} + {\gamma _{23}} - {\gamma _1} - {\gamma _2} - {\gamma _3}} \right) ,
\end{equation}
where 
\begin{equation}\label{s3.5}
{\gamma _{jl}} = \sum\limits_{k \ne N} {\frac{{{{\left| {\left\langle {{p_j}} \right|\left. {{\lambda _k}} \right\rangle  + \left\langle {{p_l}} \right|\left. {{\lambda _k}} \right\rangle } \right|}^2}}}{{{\lambda _k}}}},
\end{equation}
and 
\begin{equation}\label{s3.6}
{\gamma _j} = \sum\limits_{k \ne N} {\frac{{{{\left| {\left\langle {{p_j}} \right|\left. {{\lambda _k}} \right\rangle } \right|}^2}}}{{{\lambda _k}}}}.
\end{equation}
Since $\gamma_{jl}$ and $\gamma_j$ are all known for us, so the $\gamma$ is. And similarily 
\begin{equation}\label{s3.7}
{\beta ^2} = \frac{1}{3}\left( {\beta _{12}^2 + \beta _{13}^2 + \beta _{23}^2 - \beta _1^2 - \beta _2^2 - \beta _3^2} \right),
\end{equation}
where 
\begin{equation}\label{s3.8}
\beta _{jl}^2 = \sum\limits_{k \ne N} {\frac{{{{\left| {\left\langle {{p_j}} \right|\left. {{\lambda _k}} \right\rangle  + \left\langle {{p_l}} \right|\left. {{\lambda _k}} \right\rangle } \right|}^2}}}{{{\lambda _k}^2}}},
\end{equation}
and 
\begin{equation}\label{s3.9}
\beta _j^2 = \sum\limits_{k \ne N} {\frac{{{{\left| {\left\langle {{p_j}} \right|\left. {{\lambda _k}} \right\rangle } \right|}^2}}}{{\lambda _j^2}}}.
\end{equation}
Hence, for a general $m$, adopt the same definition in (\ref{s3.5}),(\ref{s3.6}),(\ref{s3.8}) and (\ref{s3.9}), we have 
\begin{equation} \label{s3.10}
\left\{ {\begin{array}{*{20}{c}}
	{\gamma  = \frac{1}{m}\left( {\sum\limits_{j \ne l} {{\gamma _{jl}}}  - \left( {m - 2} \right)\sum\limits_j {{\gamma _j}} } \right)}\\
	\\
	{{\beta ^2} = \frac{1}{m}\left( {\sum\limits_{j \ne l} {{\beta _{jl}^2}}  - \left( {m - 2} \right)\sum\limits_j {{\beta _j^2}} } \right)}
	\end{array}} \right.
\end{equation}
The values of $\gamma_{jl},\gamma_j,\beta_{jl}$ and $\beta_j$ can be computed by the method of Sec.3. This means that we can figure out the amplitude of $m$-vertex uniform superposition state search by using the result of a single vertex and two-vertex search.

\section{\label{sec:level1}The Optimal Condition}
In this section, we focus on the search of any state that has a bigger than or equal to $1/ \sqrt N$ overlap with the initial state. We hope to obtain a Laplacian spectrum condition of the associated graph on which we can perform optimal search for any such state. The target state can be represented as combinations of the vertex basis, 
\begin{equation}
\left| w \right\rangle  = \sum\limits_j {{w_j}\left| j \right\rangle } .
\end{equation}
It can also be represented as combinaions of Laplacian eigenvectors of the evolved graph: 
\begin{equation}
\left| w \right\rangle  = \sum\limits_j {{P_j}\left| {{\lambda _j}} \right\rangle } .
\end{equation}
where ${P_j} = \left\langle w \right|\left. {{\lambda _j}} \right\rangle $, and let $a_j=P_j^2$.
For any two Laplacain eigenvalues, we have:
\begin{equation}\label{s4.1}
\left| \frac{1}{{\lambda _j}} - \frac{1}{{\lambda _k}} \right| \le \theta,
\end{equation}
where $\theta = \frac{1}{\lambda_{N-1}}-\frac{1}{\lambda_1}$ is the difference between the reciprocal of the nonzero minimal and maximal eigenvalues, and we have assumed that the number of less than $1$ non-zero eigenvalue is at most $1$. Therefore, an arbitrary configuration $\left\{ {{a_1}, \cdots ,{a_{N}}} \right\}$ corresponds a target state. And the configuration $\left\{ {{a_1}, \cdots ,{a_{N }}} \right\}$ can be regarded as a probability distribution, 
\begin{equation}
\sum\limits_j {{a_j}}  = 1.
\end{equation}
Thus $\gamma$ can be regarded as the expectation of $\frac{1}{\lambda_k}$, hence $\gamma$ and $\beta$ can be related together by the equation of the variance formula. By taking $a_j$ in, we have the difference:
\begin{equation}\label{s4.2}
\left | \frac{1}{{\lambda _j}} - \sum\limits_{k \ne N} {\frac{{{a_k}}}{{\lambda _k}}} \right |\le \theta.
\end{equation}
Then
\begin{equation} \label{s4.3}
\sum\limits_{j \ne N} {{a_j}{{\left( {\frac{1}{{{\lambda _j}}} - \sum\limits_{k \ne N} {\frac{{{a_k}}}{{{\lambda _k}}}} } \right)}^2}} \le {\theta ^2}.
\end{equation}
Expand the left side of (\ref{s4.3}) to obtain
\begin{equation} \label{s4.4}
\sum\limits_{j \ne N} {{a_j}{{\left( {\frac{1}{{{\lambda _j}}} - \sum\limits_{k \ne N} {\frac{{{a_k}}}{{{\lambda _k}}}} } \right)}^2}} \approx \sum\limits_{j \ne N} {\frac{{{a_j}}}{{\lambda _j^2}}}  - {\left( {\sum\limits_{k \ne N} {\frac{{{a_k}}}{{{\lambda _k}}}} } \right)^2} .
\end{equation}
Hence, we have 
\begin{equation} \label{s4.5}
\sum\limits_{j \ne N} {\frac{{{a_j}}}{{\lambda _j^2}}}  - {\left( {\sum\limits_{k \ne N} {\frac{{{a_k}}}{{{\lambda _k}}}} } \right)^2} \le {\theta ^2}.
\end{equation}
The left side of (\ref{s4.5}) is similar with the definition of variance. Divide by $\sum\limits_{k \ne N} {\frac{{{a_k}}}{{\lambda _k^2}}} $ on both sides of equation (\ref{s4.5}) and shift the entry to obtain:
\begin{equation}\label{s4.6}
1 - \frac{{{\theta ^2}}}{{{\beta ^2}}}  \le \frac{{{\gamma ^2}}}{{{\beta ^2}}}.
\end{equation}
Since the optimal search require $\gamma/\beta \geq \frac{1}{\sqrt 2}$, we let the left side of (\ref{s4.6}) be larger than $1/2$. Bring $\theta$ and $\beta$ into equation (\ref{s4.6}) to obtain:
\begin{equation}\label{s4.7}
{\left( {{\lambda _1} - {\lambda _{N - 1}}} \right)^2} \le \frac{1}{2}\lambda _1^2\lambda _{N - 1}^2\sum\limits_{k \ne N} {\frac{{{a_k}}}{{\lambda _k^2}}}. 
\end{equation}
Once for arbitrary state, the inequality (\ref{s4.7}) hold, that is to say, for arbitrary configuration $\left\{ {{a_1}, \cdots ,{a_{N - 1}}} \right\}$ the  (\ref{s4.7}) satisfied. This leads to
\begin{equation}\label{s4.8}
\frac{{{{\left( {{\lambda _1} - {\lambda _{N - 1}}} \right)}^2}}}{{\lambda _{N - 1}^2}} \le \frac{1}{2}.
\end{equation}
Therefore as if (\ref{s4.8}) satisfies, then on that graph, the search is optimal for any marked state. We simplify expression (\ref{s4.8}) to obtain
\begin{equation} \label{s4.9}
\frac{\lambda _1}{\lambda _{N - 1}} \le 1 + \frac{1}{{\sqrt 2}}.
\end{equation}
Where $\lambda_1$ and $\lambda_{N-1}$ are the maximum and minimum non-zero eigenvalue of the Laplacian matrix respectively. This is different from the single vertex state search in which only the energy gap between the maximum and the second maximum eigenvaue is confined\cite{childs2004spatial}. Since, the (\ref{s4.9})
doesn't contain any entry about eigenvectors, one can judge, only need to know its spectrum, if a graph is suitable for performing spatial search of a general state. And we will give three examples in the next section.

\section{\label{sec:level1}Three kinds of graphs}
In the previous section, we obtain a Laplacian spectrum condition. Graphs that hold the condition can be used to do optimal search for general state. That condition can help us determine some special graph although we don't know the exact eigenspace of the graph Laplacian matrix. 

\textit{Induced graph from the complete graph}.
Obviously, the complete graphs satisfy that condition. We can obtain new type of graphs which also satisfy the condition by induced from complete graph.
Let $G$ be the graph induced from the complete graph by deleting $l$ disjoint edges, where $2l\le n$. From \cite{bapat2010graphs} we know that the eigenvalues of $G$ are
$n$,$n-2$,$0$, the corresponding multiplicity are $n-l-1$,$l$,and $1$. Thus for the graph $G$ we have 
\begin{equation} \label{s5.1}
\frac{\lambda_1 - \lambda_{N-1}}{\lambda_{N-1}} = \frac{2}{N-2}.
\end{equation}
Therefore, when the vertex number of the original complete graph is larger than 4, then, the induced graph $G$ satisfies the expression (\ref{s4.8}).
\begin{figure}[!htbp]\label{fig:1}
	\centering\includegraphics[width=0.4\textwidth]{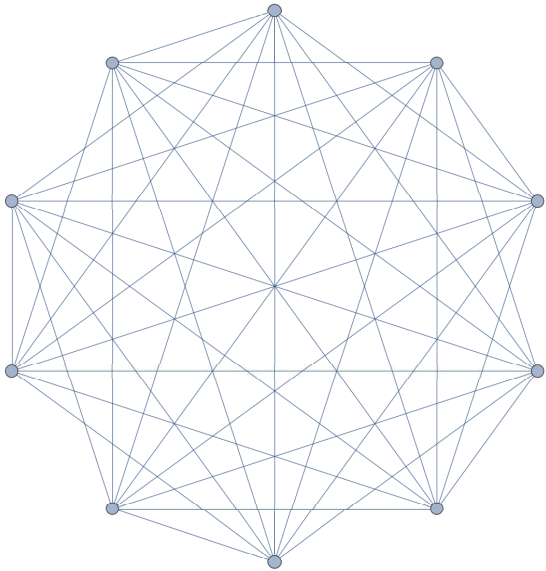}
	\caption{An induced graph by deleting 5 edges in the $K_{10}$}
\end{figure}

\indent
\textit{Strongly regular graph}.
The strongly regular graphs $(SRG)$ can be utilized to do optimal search for one single vertex by CTQW \cite{janmark2014global}. And since the adjacent spectrum is known to us\cite{ChrisGodsil2004Algebraic}, so as the Laplacian spectrum.

\begin{figure}[htbp]\label{fig:2}
	\centering\includegraphics[width=0.4\textwidth]{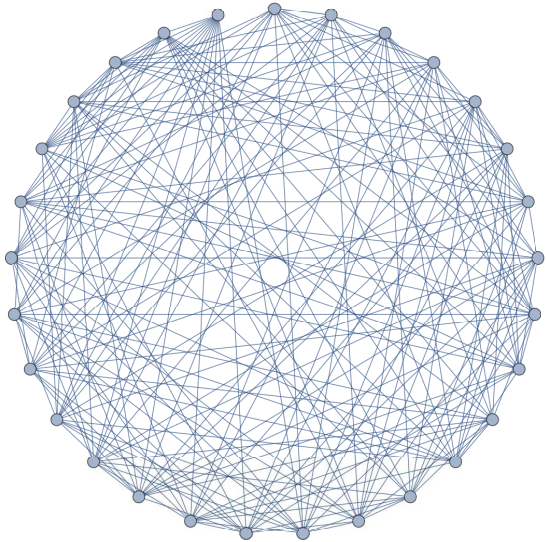}
	\caption{The strongly regular graph with parameters (29,14,6,7)}
\end{figure}
A strongly regular graph has parameters $(n,k,a,c)$, let $\Delta {\rm{ = }}{\left( {a - c} \right)^2} + 4\left( {k - c} \right)$, then its adjacent eigenvalues are $k$ and $\frac{1}{2} (a-c\pm \sqrt \Delta)$. And, its Laplacian eigenvalues are $k-\frac{1}{2} (a-c\pm \sqrt \Delta)$ and $0$. Hence, we have 
\begin{equation} \label{s5.2}
\frac{\lambda_1 - \lambda_{N-1}}{\lambda_{N-1}} = \frac{{\sqrt \Delta  }}{{\left( {k - \frac{1}{2}(a - c + \sqrt \Delta  )} \right)}}.
\end{equation}
When we bring the parameters of the $SRG$ into (\ref{s5.2}) then we will immediately know whether the (\ref{s4.8}) satisfies or not. Let us consider the $SRG$ with parameters $(29,14,6,7)$, we find that this $SRG$ satisfy the (\ref{s4.8}). So, one can do optimal search for any state on $SRG$ with $(29,14,6,7)$.

\indent
\textit{Regular complete multi-partite graph}. We consider the complete m-partite graph in which there is an edge between every pair of vertices from different independent sets. We let the number of vertices in each independent set is a constant $k$. Then the graph becomes a regular complete m-partite graph.
\begin{figure}[!htbp]\label{fig:3}
	\centering\includegraphics[width=0.4\textwidth]{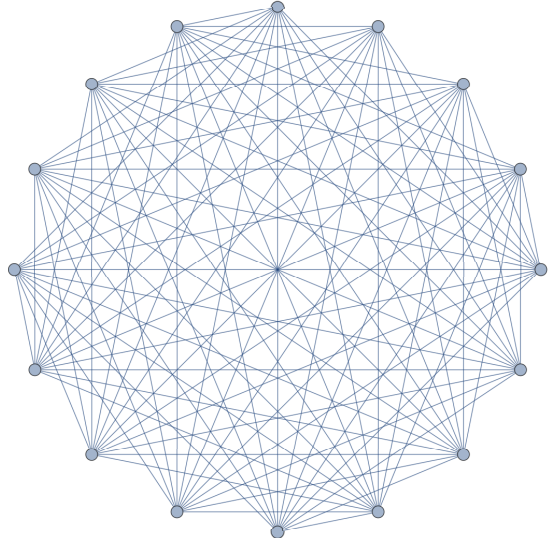}
	\caption{The regular complete 4-partite graph, the vertex number in every independent set is 4, its Laplacian eigenvaues are $16$, $12$ and 0}
\end{figure}
The adjacent matrix of the regular complete m-partite graph are 
\begin{equation}
{A_{m - partite}} = \left[ {\begin{array}{*{20}{c}}
	{{O_k}}&{{J_{k \times k}}}& \cdots &{{J_{k \times k}}}\\
	{{J_{k \times k}}}&{{O_k}}& \cdots &{{J_{k \times k}}}\\
	\vdots &{}& \vdots & \vdots \\
	{{J_{k \times k}}}& \cdots & \cdots &{{O_k}}
	\end{array}} \right].
\end{equation}
In this case, we have $A_{m-partite} = A_{K_m}\otimes J_{k\times k}$, the adjacent spectrum of $A_{m-partite}$ 
is $\{(m-1)k,-k,0 \}$. Since, the graph is regular, then its Laplacian spectrum is $\{mk,(m-1)k,0\}$. We have 
\begin{equation} \label{s5.3}
\frac{(\lambda_1-\lambda_{N-1})^2}{ \lambda_{N-1}^2}=\frac{1}{(m-1)^2}.
\end{equation}
So if we let $m>2$ then the spectrum of graph satisfies (\ref{s4.8}), hence it is optimal.

\section{\label{sec:level1}Conclusion}
We have examined the multiple solutions search problem with the spatial search by CTQW. We have computed the search amplitude of two-vertex uniform superposition state on hypercube graph. Since most of the graphs can be used to do optimal search only for partial states in the whole state space, we need to derived the condition of graphs on which the spatial search by CTQW are optimal for all such states which are not orthogonal to the initial state. And we have applied the condition to three examples, the induced complete graph, the strongly regular graph and the regular complete muti-partite graphs, with proper configuration of their graph parameters, one can do optimal search on these graphs.

\indent

\textbf{Acknowledgements}
This work is supported by the National Natural Science Foundation of China (61502101), the Natural Science Foundation of Jiangsu Province, China (BK20140651)
, the Natural Science Foundation of Anhui Province, China (Grant No.1708085MF162) and Natural Science Foundation of Jiangsu Province (Grant No. BK20171458) .

\section{APPENDIX A}
In this section, we provide the procedures of deriving the amplitude of the marked state. The methods and procedures are similar to those reported in [4]
We rewrite the eigenequation as follow:
\begin{equation}\label{s1}
H\left| {{\mu _\kappa }} \right\rangle  = {\mu _\kappa }\left| {{\mu _\kappa }} \right\rangle,
\end{equation}
where 
\begin{equation}\label{a1}
H =  \gamma L - \left| w \right\rangle \left\langle w \right|.
\end{equation}
Define $ {R_k} = {\left| {\left\langle {w\left| {{\mu _k}} \right\rangle } \right.} \right|^2}$, and bring (\ref{a1}) into equation into (\ref{s1}) and 
\begin{equation}\label{s2}
\left| {{\mu _k}} \right\rangle  = \sqrt {{R_k}} {\left( {\gamma L - {\mu _k}} \right)^{ - 1}}\left| w \right\rangle.
\end{equation}
Multiply by $\left| w \right\rangle$ on the left, results in:
\begin{equation}\label{s3}
\left\langle w \right|{\left( {\gamma L - {\mu _k}} \right)^{ - 1}}\left| w \right\rangle  = 1.
\end{equation}
The eigenvectors $\left\{ {\left| {{\lambda _1}} \right\rangle , \ldots ,\left| {{\lambda _n}} \right\rangle } \right\}$ of $L$ constitute a group of standard orthogonal basis; then the marked state $\left| w \right\rangle $ has a unique representation as
\begin{equation}\label{s4}
\left| w \right\rangle  = \sum\limits_j {\left| {{\lambda _j}} \right\rangle } \left\langle {{\lambda _j}} \right|\left. w \right\rangle  = \sum\limits_j {P_j^* \left| {{\lambda _j}} \right\rangle },
\end{equation}
where ${P_i} = \left\langle w \right|\left. {{\lambda _j}} \right\rangle $. By combining equations  (\ref{s3}) and (\ref{s4}), we obtain
\begin{equation}\label{s5}
\left\langle w \right|{\left( { \gamma L - {\mu _k}} \right)^{ - 1}}\left| w \right\rangle  = \sum\limits_j {\frac{{P_j^2}}{{\gamma {\lambda _j} - {\mu _k}}}}.
\end{equation}
Here define:
\begin{equation}\label{s6}
f\left( \mu  \right) = \sum\limits_j {\frac{{P_j^2}}{{\gamma {\lambda _j} - \mu }}},
\end{equation}
then $f\left( {{\mu _k}} \right) = 1$, Since
\begin{equation}\label{s7}
\left\langle {{\mu _k}} \right.\left| {{\mu _k}} \right\rangle  = {R_k}\left\langle w \right|{\left( { \gamma L - {\mu _k}} \right)^{ - 2}}\left| w \right\rangle  = 1.
\end{equation}
Take equation (\ref{s6}) into equation (\ref{s7})
\begin{equation}\label{s8}
\left\langle {{\mu _k}} \right.\left| {{\mu _k}} \right\rangle  = {R_i}\left\langle w \right|{\left( { \gamma L - {\mu _i}} \right)^{ - 2}}\left| w \right\rangle .
\end{equation}
One calculate (\ref{s8}) to obtain
\begin{equation}\label{s8}
\left\langle {{\mu _k}} \right.\left| {{\mu _k}} \right\rangle={R_i}\sum\limits_i {\frac{{{{\left| {{P_i}} \right|}^2}}}{{{{\left( {\gamma {\lambda _i} - {\mu _k}} \right)}^2}}}}.
\end{equation}
Therefore ${R_k} = \frac{1}{{{f^\prime}(\mu )}}$ and since the initial state $\left| s \right\rangle$ is one of the eigenvector of $L$ with eigenvalue 0, we have
\begin{equation}\label{s9}
\left\langle s \right.\left| {{\mu _k}} \right\rangle = \sqrt {{R_k}} \left\langle s \right|{\left( {\gamma L - {\mu _k}} \right)^{ - 1}}\left| w \right\rangle .
\end{equation}
This leads to 
\begin{equation}\label{s9}
\left\langle s \right.\left| {{\mu _k}} \right\rangle = -P_N \frac{{\sqrt {{R_k}} }}{{\mu _k}}.
\end{equation}
At time $t$, the amplitude of marked states is
\begin{equation}\label{s10}
\left\langle w \right|\mathop e\nolimits^{ - iHt} \left| s \right\rangle  = \sum\limits_k {\left\langle w \right.\left| {{\mu _k}} \right\rangle } \left\langle {{\mu _k}} \right.\left| s \right\rangle \mathop e\nolimits^{ - i{\mu _k}t}.
\end{equation}
We bring all the results into equation (\ref{s10}) to obtain the amplitude equation:
\begin{equation}\label{s11}
\left\langle w \right|\mathop e\nolimits^{ - iHt} \left| s \right\rangle  =  - P_N \sum\limits_k {\frac{{\mathop e\nolimits^{ - i{\mu _k}t} }}{{{\mu _k}{f^\prime}({\mu _k})}}}.
\end{equation}
Separating equation (\ref{s6}) into the sum of two parts results in:
\begin{equation}
\label{s1.20}
f\left(\mu\right) =-\frac{{P_N^2}}{\mu}+\sum\limits_{k\ne N}{\frac{{P_k^2}}{{\gamma {\lambda _k}-\mu }}}.
\end{equation}
If $\left| \mu  \right| \ll \gamma {\lambda _i}$, based on the Taylor expansion, we have:
\begin{equation}
\label{s1.21}
f\left(\mu\right)\approx-\frac{{P_N^2}}{\mu } + \frac{1}{\gamma }\sum\limits_{k\ne N}{\frac{{P_k^2}}{{{\lambda _k}}}}+\frac{\mu}{{{\gamma ^2}}}\sum\limits_{k\ne N}{\frac{{P_k^2}}{{\lambda_k^2}}}.
\end{equation}
Setting $\gamma  = \sum\limits_{k \ne N} {\frac{1}{{{\lambda _k}}}P_k^2}$ , and let (\ref{s1.21}) equal to 1, the two eigenvalues can be solved as:
\begin{equation}
\label{s1.22}
\left\{ {\begin{array}{*{20}{c}}
	{{\mu _1} = \frac{{\gamma {P_N}}}{{\sqrt {\sum\limits_{k \ne N} {\frac{{P_i^2}}{{\lambda _k^2}}} } }}}\\
	\\
	{{\mu _2} = \frac{{ - \gamma {P_N}}}{{\sqrt {\sum\limits_{k \ne N} {\frac{{P_i^2}}{{\lambda _k^2}}} } }}}
	\end{array}} \right.
\end{equation}
From (\ref{s1.21}), we have
\begin{equation}
\label{s1.23}
f^\prime\left(\mu\right)\approx\frac{{P_N^2}}{{{\mu^2}}}+\frac{1}{{{\gamma^2}}}\sum\limits_{k \ne N} {\frac{{P_i^2}}{{\lambda _k^2}}}.
\end{equation}
Substituted equation  (\ref{s1.22}) into equation (\ref{s1.23}) results in:
\begin{equation}
\label{s1.24}
{f^\prime}\left( {{\mu _1}} \right) \approx {f^\prime}\left( {{\mu _2}} \right) \approx \frac{2}{{{\gamma ^2}}}\sum\limits_{i \ne N} {\frac{{P_k^2}}{{\lambda _k^2}}}.
\end{equation}
When $t=0$, the result of equation (\ref{s11}) is $P_N$, therefore, the sum of all entries except the first two in equation in (\ref{s11}) is
\begin{equation}
\label{s1.25}
- P_N\sum\limits_{i > 2} {\frac{1}{{{\mu _i}{f^\prime}({\mu _k})}}}  = P_N \left( {1 + \frac{1}{{{\mu _1}{f^\prime}\left( {{\mu _1}} \right)}} + \frac{1}{{{\mu _2}{f^\prime}\left( {{\mu _2}} \right)}}} \right).
\end{equation}
Since ${\mu _1}{f^\prime}\left( {{\mu _1}} \right) =  - {\mu _2}{f^\prime}\left( {{\mu _2}} \right)$, therefore, the contribution of the entries greater than $k = 2$ is far less than 1, we ignored them so that the result of equation (\ref{s11}) approximates:
\begin{equation}\label{s1.26}
\left| {\left\langle w \right|\mathop e\nolimits^{ - iHt} \left| s \right\rangle } \right|  \approx P_N \left| {\frac{{2\sin \left( {{\mu _1}t} \right)}}{{{\mu _1}{f^\prime}\left( {{\mu _1}} \right)}}} \right| .
\end{equation}
As the definition of $\gamma$ and $\beta$, the (\ref{s1.26}) can be written as 
\begin{equation}
\left| {\left\langle w \right|\mathop e\nolimits^{ - iHt} \left| s \right\rangle } \right|  \approx \frac{\gamma }{\beta }\left| {\sin  \left(  {\frac{\gamma P_N  }{{\beta }}t}  \right)} \right|.
\end{equation}

\indent


\begin{thebibliography}{10}
	
	\bibitem{grover1997quantum}
	Lov~K Grover.
	\newblock Quantum mechanics helps in searching for a needle in a haystack.
	\newblock {\em Physical review letters}, 79(2):325, 1997.
	
	\bibitem{zalka1999grover}
	Christof Zalka.
	\newblock Grover¡¯s quantum searching algorithm is optimal.
	\newblock {\em Physical Review A}, 60(4):2746, 1999.
	
	\bibitem{Cerf1998Nested}
	N.~J. Cerf, L.~K. Grover, and C.~P. Williams.
	\newblock Nested quantum search and np-complete problems.
	\newblock 8(3):453--474, 1998.
	
	\bibitem{Roland2003Adiabatic}
	Jérémie Roland and Nicolas~J. Cerf.
	\newblock Adiabatic quantum search algorithm for structured problems.
	\newblock {\em Physical Review A}, 68(6):150--150, 2003.
	
	\bibitem{Albash2018Adiabatic}
	Tameem Albash and Daniel~A. Lidar.
	\newblock Adiabatic quantum computation.
	\newblock {\em Rev.mod.phys}, 90(1), 2018.
	
	\bibitem{Farhi1998Quantum}
	Edward Farhi and Sam Gutmann.
	\newblock Quantum computation and decision trees.
	\newblock {\em Phys.rev.a}, 58(2):915--928, 1998.
	
	\bibitem{Christandl2003Perfect}
	M~Christandl, N~Datta, A~Ekert, and A.~J. Landahl.
	\newblock Perfect state transfer in quantum spin networks.
	\newblock {\em Physical Review Letters}, 92(18):187902, 2003.
	
	\bibitem{M2011Continuous}
	Oliver Mülken and Alexander Blumen.
	\newblock Continuous-time quantum walks: Models for coherent transport on
	complex networks.
	\newblock {\em Physics Reports}, 502(2-3):37--87, 2011.
	
	\bibitem{Kulvelis2015Universality}
	N~Kulvelis, M~Dolgushev, and O~Mülken.
	\newblock Universality at breakdown of quantum transport on complex networks.
	\newblock {\em Physical Review Letters}, 115(12):120602, 2015.
	
	\bibitem{Galiceanu2016dendrimer}
	Mircea Galiceanu and Walter~T. Strunz.
	\newblock Continuous-time quantum walks on multilayer dendrimer networks.
	\newblock {\em Phys. Rev. E}, 94:022307, Aug 2016.
	
	\bibitem{Qiang2012An}
	Xiaogang Qiang, Xuejun Yang, Junjie Wu, and Xuan Zhu.
	\newblock An enhanced classical approach to graph isomorphism using
	continuous-time quantum walk.
	\newblock {\em Journal of Physics A Mathematical General}, 45(4):119--156,
	2012.
	
	\bibitem{Wong2015Grover}
	Thomas~G. Wong.
	\newblock Grover search with lackadaisical quantum walks.
	\newblock {\em Journal of Physics A Mathematical \& Theoretical},
	48(43):435304, 2015.
	
	\bibitem{Gamble2010Two}
	John~King Gamble, Mark Friesen, Dong Zhou, Robert Joynt, and S.~N. Coppersmith.
	\newblock Two-particle quantum walks applied to the graph isomorphism problem.
	\newblock {\em Physical Review A}, 81(5):90--90, 2010.
	
	\bibitem{Childs2009Universal}
	A.~M. Childs.
	\newblock Universal computation by quantum walk.
	\newblock {\em Physical Review Letters}, 102(18):180501, 2009.
	
	\bibitem{childs2004spatial}
	Andrew~M Childs and Jeffrey Goldstone.
	\newblock Spatial search by quantum walk.
	\newblock {\em Physical Review A}, 70(2):022314, 2004.
	
	\bibitem{janmark2014global}
	Jonatan Janmark, David~A Meyer, and Thomas~G Wong.
	\newblock Global symmetry is unnecessary for fast quantum search.
	\newblock {\em Physical Review Letters}, 112(21):210502, 2014.
	
	\bibitem{chakraborty2016spatial}
	Shantanav Chakraborty, Leonardo Novo, Andris Ambainis, and Yasser Omar.
	\newblock Spatial search by quantum walk is optimal for almost all graphs.
	\newblock {\em Physical review letters}, 116(10):100501, 2016.
	
	\bibitem{Nielsen2011Quantum}
	Michael~A. Nielsen and Isaac~L. Chuang.
	\newblock {\em Quantum Computation and Quantum Information: 10th Anniversary
		Edition}.
	\newblock Cambridge University Press, 2011.
	
	\bibitem{Williams2011Explorations}
	Colin~P. Williams.
	\newblock {\em Explorations in quantum computing}.
	\newblock Springer London, 2011.
	
	\bibitem{Wong2016Spatial}	
	Wong, Thomas G.
	\newblock {\em Spatial search by continuous-time quantum walk with multiple marked vertices}.
	\newblock {Quantum Information Processing}, 15(4):1411--1443, 2016
	
	\bibitem{Fiedler1973Algebraic}
	Miroslav Fiedler.
	\newblock Algebraic connectivity of graphs.
	\newblock {\em Czechoslovak Mathematical Journal}, 23(23):298--305, 1973.
	
	\bibitem{merris1994laplacian}
	Russell Merris.
	\newblock Laplacian matrices of graphs: a survey.
	\newblock {\em Linear algebra and its applications}, 197:143--176, 1994.
	
	\bibitem{Mieghem2011Graph}
	Piet~Van Mieghem.
	\newblock {\em Graph Spectra for Complex Networks}.
	\newblock Cambridge University Press, 2011.
	
	\bibitem{HypergeometricPFQ}
	Wolfram.
	\newblock {HypergeometricPFQ}.
	\newblock
	\url{http://reference.wolfram.com/language/ref/HypergeometricPFQ.html}, 1999.
	\newblock [Online; accessed 17-October-2017].
	
	\bibitem{bapat2010graphs}
	Ravindra~B Bapat.
	\newblock {\em Graphs and matrices}, volume~27.
	\newblock Springer, 2010.
	
	\bibitem{ChrisGodsil2004Algebraic}
	Chris Godsil and Gordon~F. Royle.
	\newblock {\em Algebraic graph theory}.
	\newblock Springer Science \& Business Media, 2013.

\end{thebibliography}
\end{document}